\documentclass[rmp,twocolumn,floatfix]{revtex4}

\usepackage{dcolumn,graphicx,amsmath,amssymb,txfonts}

\begin{document}

\title{Currency and commodity metabolites:\\ Their identification and
  relation to the modularity of metabolic networks}

\author{Mikael Huss}
\affiliation{School of Computer Science and Communication,
  Royal Institute of Technology, 100 44 Stockholm, Sweden}

\author{Petter Holme}
\affiliation{Department of Computer Science, University of New Mexico,
  Albuquerque, NM 87131, U.S.A.}

\begin{abstract}
  The large-scale shape and function of metabolic networks are
  intriguing topics of systems biology. Such networks are on one hand
  commonly regarded as modular (i.e.\ built by a number of relatively
  independent subsystems), but on the other hand they are robust in a way
  not expected of a purely modular system. To address this question
  we carefully discuss the partition of metabolic networks into
  subnetworks. The practice of preprocessing such networks by removing
  the most abundant substrates, ``currency metabolites,'' is
  formalized into a network-based algorithm. We study partitions
  for metabolic networks of many organisms and find cores of currency
  metabolites and modular peripheries of what we call ``commodity
  metabolites.'' The networks are found to be more modular than random
  networks but far from perfectly divisible into modules. We argue
  that cross-modular edges are the key for the robustness of
  metabolism.
\end{abstract}

\maketitle

\section{Introduction}

For well over half a century, metabolism has been described as modular,
i.e.\ divisible into relatively autonomous subunits such
as the citric acid cycle~\cite{citric}, glycolysis~\cite{stryer:bio},
etc. For the pioneers of the 20th century it was a great feat to
describe such connected set of reactions. But do these subunits tell
us that the organization of metabolic networks is fundamentally
modular, or are they a result of the limited knowledge of the time
they were discovered? 

For systems, such as the metabolism, where a continuous flow is needed
throughout a large part of the system, a modular organization of a
system is less robust than a more integrated
topology~\cite{wagner:robu}---and metabolic networks are robust;
metabolite fluxes are restored in minutes after large
perturbations~\cite{palsson:minutes}. A very modular system would be
expected to consist of modules with simple (or narrow) inputs and
outputs and a more complex interior. Severing the input of a module
(in a biochemical context, this could happen through mutation or
exposure to atypical conditions, e.g. different types of starvation)
would then affect the whole functioning of the module. On the other
hand, a homogeneous network without explicit interfaces between
modules would be very robust in a way termed ``distributed
robustness''~\cite{wagner:robu}. With the biochemical reaction data
available today, answers to questions around modularity and robustness
of cellular systems is within reach, and we will argue that
biochemical networks are not modular or distributed, but probably best
described as having a little bit of both.

To address such general questions one soon gets into
technicalities. First of all, one has to choose an appropriate level
of description. As many other large-scale studies~\cite{zhao:meta} we
simplify the biochemistry to a network and use graph theory to
describe its organization. Here we use a \textit{substrate graph}
representation where chemical compounds are nodes and (undirected)
edges connect nodes if one of them can produce the other through a
reaction; other representations (reaction graphs, enzyme-centric
graphs and bipartite compound-reaction graphs) have been used (see
Ref.~\cite{zhao:meta} and references therein). Arita~\cite{arita:not}
introduced yet another representation based on the carbon atoms that
are actually transferred during metabolic reactions, and argued that
other representations give a skewed estimate of average path
lengths. However, he also stated that his structure-based description
in a sense gives a compressed view of metabolism, making it difficult
to assess the overall robustness of the networks~\cite{arita:not},
which is one of our aims in this study. 

Although biochemical modules are ultimately dynamical
entities~\cite{dyn:mod} there is a prevailing supposition that the
modules of metabolism can be identified with network clusters (densely
connected regions of the
networks)~\cite{our:snwkh,ravasz:hier,gui:meta,ma:meta,ma:zeng}. But
identifying clusters is far from straightforward. One has to choose
network representation, cluster-detection algorithm and, last but not
least, whether or not to preprocess the data by removing abundant
metabolites. The logic behind such preprocessing is that such
substrates (like water, carbon dioxide or adenosine triphosphate) are
so plentiful in normally functioning cells that their concentrations
put no constraints on the activity of a module; in other words, they
can be regarded as externally buffered with respect to the
system~\cite{schus:dec}. Such ubiquitous substrates are sometimes, by
analogy to economy, termed \textit{currency metabolites}---they have a
high turnover and occur in widely different exchange processes. (In
some studies they are instead referred to as ``current'' metabolites,
emphasizing their flow through the metabolic networks.) For example,
adenosine triphosphate (ATP) can be seen as the energy currency of the
cell. Continuing this analogy we will call non-currency metabolites
\textit{commodity metabolites}. So far, the identification of currency
metabolites has made based on non-formalized, chemical
considerations. Among authors who have chosen to preprocess their
metabolic network data some compounds (ATP and NADH, for example) are
virtually always considered currency metabolites, others (e.g.\ small
molecules like water and oxygen) are sometimes removed (e.g.\
Ref.~\cite{ma:zeng}) and sometimes not (e.g.\
Ref.~\cite{wagner:sw}). Since currency metabolites have high degrees
(number of neighbors),
they turn up as ``hub metabolites'' in studies where abundant
substances have not been filtered away (e.g.\
Ref.~\cite{jeong:meta}. Interestingly, versatile currency or hub
metabolites like ATP, NADH and $\textrm{H}_2\textrm{O}$  seem to
support enzyme variability and stimulate pathway
evolution~\cite{schmidt:metab}. We propose a way to identify a set of
currency metabolites from the network structure alone, and, in
extension, a scheme to study the modularity of metabolism that does
not rely on outer information about the substrates. With this
carefully justified network decomposition scheme we look at the
large-scale organization of metabolic networks from several different
organisms.

In the rest of this paper we will describe how the networks are
compiled, state the precise definition of the algorithm, and present
results from the analysis of a number of organisms.

\section{Network construction}

Interaction networks were computed using data downloaded (between
March 13 and  26, 2006) from the anonymous FTP service of KEGG (Kyoto
Encyclopedia of Genes and Genomes) at
\url{http://www.genome.jp/anonftp/}. The following steps were used for
each organism (109 in all): 1. A list of the known enzymes encoded by
the organism's genome was downloaded from the ``genomes''
database of KEGG. 2. The file specifying reactions in the ``ligand''
database of KEGG was scanned for all reactions catalyzed by enzymes
present in the organism in question. 3. reactants and products were
extracted for each of these reactions and all resulting
reactant-product pairs were written to an edge-list specifying the
connections between the substances. A set of Python scripts automated
the process so that all files could be generated in a single batch
without manual intervention. In the resulting networks for each
organism, substances become nodes and the links between substances
from the link file become edges; we will call the number of nodes $N$
and the number of edges $M$ in the following.  The adjacency matrix
$A$ for a graph is defined as a matrix where the element $A_{ij}$ is
set to one if the edge $(i,j)$ exists in the graph and zero if it does
not.

\section{Modules and network clusters}

The problem to divide a network into subnetworks that are relatively
densely connected within, and sparsely inter-connected is an old graph
theoretical problem that recently has experienced a second
blooming. Part of the difficulty to construct such \textit{graph
  clustering} algorithms is that the objective is not completely
well-defined---the proper definition of a densely-connected
cluster is, to some extent, problem dependent. One definition
commonly used by many modern clustering
algorithms~\cite{gui:mod,arenas:clu,mejn:spectrum} is 
\begin{equation}\label{eq:q}
  Q=\sum_i\left[e_{ii}-\left(\sum_je_{ij}\right)^2\right],
\end{equation}
where the sum is over a partition into clusters and $e_{ij}$ is
the fraction of edges that leads between vertices of cluster $i$ and
$j$. Given a partition of a network into clusters $Q$ is the fraction
of vertices within clusters minus the expected fraction of edges if
the edges are wired with no structural bias. The division of networks
that maximize $Q$ is then usually taken as the desired
partition. Sampling all divisions is clearly infeasible---different
algorithms typically differs by their methods to perform this
sampling. The specific algorithm we use, presented in
Ref.~\cite{mejn:spectrum} is based one a spectral method that, roughly
speaking,  iteratively splits clusters until a further split would
increase $Q$ (for technical details, see
Ref.~\cite{mejn:spectrum}). More precisely, for a subgraph $H$ define
a matrix $\mathbf{B}$ with elements 
\begin{equation}\label{eq:b}
  B_{ij}=A_{ij}-\frac{k_ik_j}{2M}-\delta_{ij}\left[h_i -
  \frac{k_i}{2M} \sum_{l\in H}k_l\right]
\end{equation}
where $h_i$ is $i$'s number of neighbors in $H$, $k_i$ is $i$'s number
of neighbors in $G$, and $\delta_{ij}$ is Kronecker's delta. Then if
$\mathbf{B}$ has a positive leading eigenvalue there is a division of
$H$ such that $Q$ increases. The division is given by the signs of the
leading eigenvector.

$Q$ is a measure of modularity with respect to different partitions of
the same graph. The maximal $Q$ value obtained during the partition
process, $\hat{Q}$, is a crude measure the modularity of a whole
graph. However, fluctuations make $\hat{Q}$ positive even for random
networks in the $N\rightarrow\infty$ limit~\cite{gui:mod}. Indeed,
finite structureless random graphs can have almost any value of
$\hat{Q}$. Instead of measuring $\hat{Q}$ we measure the difference
between $\hat{Q}$ and $\hat{Q}$ averaged over an ensemble of
null-model networks. Since the degree of a vertex is a rather
intrinsic quantity, related to molecular traits (and the set of other
present substrates) we choose random graphs conditioned on the set of
degrees as our null-model. Such null-model networks can be
instantiated by randomly rewiring the original
network~\cite{roberts:mcmc}.

\begin{figure}
  \resizebox*{0.95 \linewidth}{!}{\includegraphics{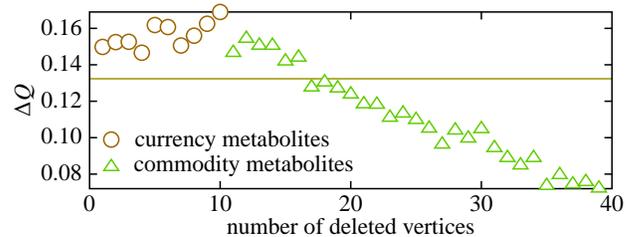}}
  \caption{ Subsequent values of the effective modularity $\Delta Q$
    during the run of the algorithm for the human metabolic
  network. The horizontal line marks the $\Delta Q$ value of the
  original network. The identified currency metabolites are (in the
  order of deletion, from left to right): water, oxygen, hydrogen ion,
  nicotinamide adenine dinucleotide phosphate (reduced form, NADPH),
  adenosine triphosphate (ATP), nicotinamide adenine dinucleotide
  phosphate (NADP), nicotinamide adenine dinucleotide (NAD$^{+}$),
  nicotinamide adenine dinucleotide (reduced form, NADH), phosphate
  and adenosine diphosphate (ADP). 100 averages of the null-model
  networks are used for the calculation of $\Delta Q$. Errorbars
  would be smaller than the symbol size.}
  \label{fig:human}
\end{figure}

\begin{figure*}
  \resizebox*{\linewidth}{!}{\includegraphics{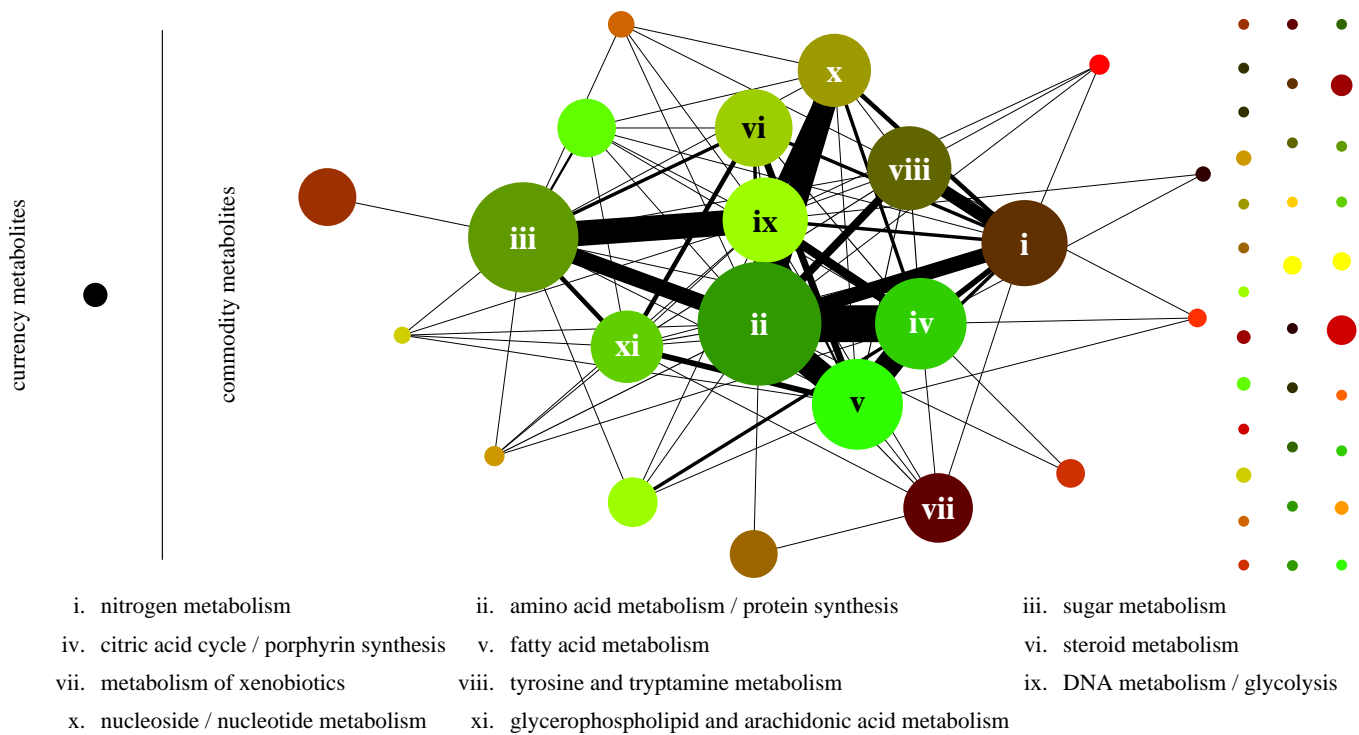}}
  \caption{ Relative sizes of the detected clusters in the human
    metabolic networks. Lines mark the connections between
    clusters. The widths of the lines are proportional to the number
    of connections. The functional assignments are done by
    inspection---they should not be viewed as absolute: a given
    assignment reflects the most common function in a cluster, but
    other functions are usually also represented to a lesser degree.}
  \label{fig:size}
\end{figure*}

How can one identify currency metabolites from a metabolic network? As
previously mentioned, currency metabolites are abundant; not only are
they present in relatively high concentrations throughout the cell,
they are also present in many different reactions. This reflects the
fact that they are used by many different enzyme
superfamilies~\cite{schmidt:metab}. That a vertex is present in many
reactions means two things: First, it has to have large
degree. Second, it will have functionally different substrates
as neighbors in the network, i.e.\ it will have edges to vertices in
different modules. The second statement means that the effective
modularity will increase when a currency metabolite is deleted. To
combine these two precepts, we delete vertices in order of degree and
take the network with the highest value of the effective modularity as
our set of commodity metabolites. More precisely, let $G_t$ be the
network after $t$ vertices are deleted, then the algorithm is as
follows:
\begin{enumerate}
\item Let $G_t$ be $G_{t-1}$ without its vertex of highest degree
  and all its incident edges. (If more than one vertex of highest
  degree exist, select one randomly.)
\item Run the clustering scheme for the current network $G_t$ and $n$
  randomizations of $G_t$.
\item Calculate the effective modularity $\Delta Q = \hat{Q} -
  \langle\hat{Q}\rangle$. If it is higher than the currently highest
  value, then save the partition.
\end{enumerate}
In practice $\Delta Q$ reaches its maximal value after about ten
iterations, seemingly independent of the network size, so the running
time of the algorithm will be a constant factor times the  running
time of the clustering algorithm ($O(N^2\log N)$ in our case).

\section{Numerical results}

\subsection{The human metabolic network: a case study}

\begin{table*}
\begin{ruledtabular}
\begin{tabular}{r|llllll}
taxonomy & number & $\langle N\rangle$ & $\langle M\rangle$ & $\langle
 N_\mathrm{currency}\rangle$ & $\langle\Delta Q_\mathrm{max}\rangle$ &
 $\langle\Delta Q_0\rangle$ \\ \hline
 animals & 5 & $1621\pm 124$&$4662\pm 473$ & $6.2 \pm 1.9$ & $0.157\pm
 0.006$ & $0.136\pm 0.002$\\
 plants & 1 & 1561 & 4302 & 1 & 0.144 & 0.130\\
 fungi & 2 & $1281\pm 97$ & $3654\pm 289$& $1.5\pm 0.5$ & $0.150\pm 0.004$
 & $0.135\pm 0.007$\\
 bacteria & 99 & $1059\pm 35$& $2739\pm 108$ & $1.7\pm 0.2$ &
 $0.140\pm 0.001$ & $0.132\pm 0.001$\\
  \end{tabular}
  \end{ruledtabular}
\vspace{4mm}
  \caption{ Statistics for different classes of organisms. The average
    number of substrates $\langle N\rangle$, edges $\langle
    M\rangle$, currency metabolites $\langle
    N_\mathrm{currency}\rangle$, average maximal effective modularity
    $\langle\Delta Q_\mathrm{max}\rangle$, average effective
    modularity  of the original network $\langle\Delta
    Q_\mathrm{max}\rangle$.}
  \label{tab:stat}
\end{table*}

Figure~\ref{fig:human} shows the effective modularity of the human
metabolic network as a function of the number of nodes removed as the
algorithm progresses. After ten removed vertices the effective
modularity reaches its maximum. After this point $\Delta Q$
decreases roughly monotonically---no larger increase of $\Delta Q$ is
observed even if one lets the algorithm run until no vertex
remain---so the human currency metabolites seem quite well-defined by
this procedure. Their identities correspond to some of the most
commonly used currency metabolites from previous studies: ATP,
NAD(P)(H), water, oxygen, the hydrogen ion etc. Indeed, our ten
currency metabolites are almost identical to the ten most abundant
metabolites in enzymatic reactions from all organisms in
KEGG~\cite{schmidt:metab} (the only difference is that our human
network has the hydrogen ion substituted for carbon dioxide).

The partition of nodes into groups for the human metabolic network at
its most modular stage of the deletion procedure (after ten currency
metabolites have been removed) is shown in
Fig.~\ref{fig:size}. Although the high-degree currency metabolites are
deleted (and with them 1988 edges) a large part of the network is
still connected. These cross-module edges make the network less than
perfectly modular and contribute to a distributed robustness.

Do the groups identified here correspond to biologically meaningful
subsets of compounds? It turns out that while some of the groups have
a clear biological interpretation, others seem more mixed, and
others---due to peculiarities of the algorithm---are isolated islands
of substances that would have been expected to end up in one of the
larger groups.
The largest clusters in Fig.~\ref{fig:size} correspond roughly to
amino acid metabolism and protein synthesis (ii), sugar metabolism
(iii) and citric acid cycle / porphyrin synthesis (iv). However, the groups are to some
extent overlapping; for example, although cluster (ii) is the main
amino acid metabolism cluster, three nodes representing amino acids occur
in the citric acid cycle cluster (iv), and the nitrogen metabolism
cluster (i) contains nodes relating to the synthesis of cysteine and
methionine. (It also contains many D amino acids, but these never
occur in proteins; most L amino acids are, as expected, located in the
amino acid metabolism and protein synthesis cluster).  Also, the
nucleotide metabolism seems to be distributed onto two clusters: one which
mainly deals with DNA metabolism and also contains several substances
relating to glycolysis, such as phosphoenolpyruvate and D-glucose
1-phosphate (ix), and one more clearly separated nucleoside/
nucleotide metabolism cluster (x). Comparing our partitions with previous studies, we note
that such cases of overlap and unclean clusters have been found in
previous studies~\cite{ravasz:hier,gui:meta}. Interestingly, the size
difference among the largest detected clusters is so small that the
size distribution hardly can be a power-law. If the networks would be
truly scale-free one would expect a power-law distribution of cluster
sizes. So even if the degree is power-law distributed, the metabolic
networks are (using the terminology of Ref.~\cite{tanaka:scalerich})
``scale-rich.''

\subsection{General organization of currency metabolites and modules}

A comparison of 109 different organisms (Table~\ref{tab:stat}) shows
the number of  currency metabolites is typically rather low and varies
little within each group of organisms. However, animals (human, mouse,
rat, fruit fly and \textit{Caenorhabditis elegans}) have significantly
more than the other groups which tend to have end up with only two
currency metabolites, typically ATP and water. This presumably reflects a
general increase in metabolic complexity in higher organisms; as
mentioned before, currency metabolites are linked to enzyme
variability and pathway evolution~\cite{schmidt:metab}. All computed
networks have markedly higher modularity than null (randomly rewired)
networks with the same degree distribution; even before any currency
metabolites have been removed. This suggests that a considerable
modularity is indeed present in metabolic networks, even if the
modules may be only partly mappable to cellular functions as
understood by contemporary biochemistry. Furthermore, the presence of
inter-modular edges (after deletion of the currency metabolites) and
the non-power-law cluster-size distribution of the human metabolic
network is observed for the vast majority of other organisms.

\section{Summary and conclusions}

In the present paper we propose a network-based method to partition
metabolites into functional groups. In concordance with other
works~\cite{schus:dec,ma:zeng,ma:meta} we propose a fundamental
dichotomy between currency and more specific, commodity
metabolites. We define the currency metabolites as the substrates
that, if omitted, increase the effective modularity of the
network. The effective modularity can be calculated by any modern
graph clustering algorithm~\cite{gui:mod,arenas:clu,mejn:spectrum} (we
use the one in Ref.~\cite{mejn:spectrum}). The same algorithm can be
used (on the fly, while tracing the core of currency metabolites) to
partition the specific metabolites into functional subgroups. Our
method is thus purely graph theoretical and does not rely on any
additional information about the substrates apart from their
connections. This is a rather simple network-view of the metabolism, in
contrast e.g.\ Ref.~\cite{gui:meta} first deletes a set of currency
metabolites from the network (based on chemical considerations in
Ref.~\cite{ma:zeng}), then proposes seven additional functional
categories among the remaining commodity vertices (defined by regions
in the event space of two vertex-specific network measures). One can
indeed proceed from our partitions and group vertices according to
other network properties, but more refined levels are not only more
prone to errors and incompleteness of the data, they are also closer
to the realm where a more complete, dynamical,
modeling~\cite{aaeau:system} is called for.

We find that the networks of all 109 organisms show a clearly positive
effective modularity. This supports the age-old idea of a modular
organization of the cellular biochemistry. But modular architectures
of systems like metabolism are also assumed to be
fragile~\cite{wagner:robu}, and metabolism is known to be robust. We
attribute this robustness to the fair amount of cross-modular
edges---after the removal of currency metabolites the network is still
largely connected. We note that some authors actually associate
modularity with robustness---damage may then be confined to a small
part of the network~\cite{kitano:robu}. While this may be true in some
cases (like the spread of disease in a population or pathogens in an
organism), it scarcely seems applicable to metabolism: If (assuming
that modules can be mapped onto canonical biochemical pathways) an
entire module corresponding to e.g.\ amino acid metabolism or
nucleotide metabolism were knocked out, the organism would hardly be
able to survive even though the rest of the modules had remained
unscathed. Rather than isolating pathways, it would seem to make sense
to intertwine them, so that metabolites whose standard synthesis
pathway has been disrupted can be synthesized through alternative
pathways.

The view of biochemistry that emerges from our study is that of a core
of currency metabolites and a periphery that is strongly, but not
completely, modular. We conjecture than the additional edges making
the modular description less than perfect are important for the
robustness of the biochemical pathways. Presumably the normal
activities of a cell can be well understood from the function of
modules, but to assess the robustness to abnormal conditions one needs
to consider the full network.

The degrees of currency metabolites are, by our definition, higher
than the average degree of commodity metabolites. This broad degree
distribution has been pointed out~\cite{jeong:meta} as a fundamental
organizational principle of metabolic networks. That the structure of
the network, manifested through a modular organization, is more
clearly visible when the currency metabolites are removed, suggests
that the functionality and active evolution is dependent on the
commodity metabolites. That the presence of high-degree vertices is an
inherent property of chemical reaction networks, rather than a result
of evolution, is supported by the fact that astrochemical networks
also show a broad, power-law-like degree
distribution~\cite{wagner:space}. The analogy with currency and
commodities is indeed quite apt: Given that we have a market
(metabolism) the presence of currency (metabolites) is inevitable. The
development (evolution) of the market occurs at the level of
commodities, but never in the absence of currency.

\begin{acknowledgements}
  P.H. acknowledges financial support from the Wenner-Gren
  foundations. The authors thank Todd Kaplan and Mark Newman for
  helpful comments.
\end{acknowledgements}

\end{document}